\documentclass[reprint,english,aip,apl,bibnotes]{revtex4-1}
\usepackage[T1]{fontenc}
\usepackage[latin9]{inputenc}
\usepackage{amstext}
\usepackage{graphicx}
\usepackage{amssymb}
\usepackage{multirow}
\usepackage{dcolumn}
\usepackage{color}
\usepackage{amsmath}
\usepackage{babel}
\usepackage{epstopdf}
\usepackage{natbib}
\makeatletter

\begin{document}
\title{Fabrication and Characterization of Aluminum Airbridges for Superconducting Microwave Circuits}
\author{Zijun Chen}
\affiliation{Department of Physics, University of California, Santa Barbara, California 93106-9530, USA}
\author{A. Megrant}
\affiliation{Department of Physics, University of California, Santa Barbara, California 93106-9530, USA}
\affiliation{Department of Materials, University of California, Santa Barbara, California 93106, USA}
\author{J. Kelly}
\affiliation{Department of Physics, University of California, Santa Barbara, California 93106-9530, USA}
\author{R. Barends}
\affiliation{Department of Physics, University of California, Santa Barbara, California 93106-9530, USA}
\author{J. Bochmann}
\affiliation{Department of Physics, University of California, Santa Barbara, California 93106-9530, USA}
\author{Yu Chen}
\affiliation{Department of Physics, University of California, Santa Barbara, California 93106-9530, USA}
\author{B. Chiaro}
\affiliation{Department of Physics, University of California, Santa Barbara, California 93106-9530, USA}
\author{A. Dunsworth}
\affiliation{Department of Physics, University of California, Santa Barbara, California 93106-9530, USA}
\author{E. Jeffrey}
\affiliation{Department of Physics, University of California, Santa Barbara, California 93106-9530, USA}
\author{J.Y. Mutus}
\affiliation{Department of Physics, University of California, Santa Barbara, California 93106-9530, USA}
\author{P.J.J. O'Malley}
\affiliation{Department of Physics, University of California, Santa Barbara, California 93106-9530, USA}
\author{C. Neill}
\affiliation{Department of Physics, University of California, Santa Barbara, California 93106-9530, USA}
\author{P. Roushan}
\affiliation{Department of Physics, University of California, Santa Barbara, California 93106-9530, USA}
\author{D. Sank}
\affiliation{Department of Physics, University of California, Santa Barbara, California 93106-9530, USA}
\author{A. Vainsencher}
\affiliation{Department of Physics, University of California, Santa Barbara, California 93106-9530, USA}
\author{J. Wenner}
\affiliation{Department of Physics, University of California, Santa Barbara, California 93106-9530, USA}
\author{T.C. White}
\affiliation{Department of Physics, University of California, Santa Barbara, California 93106-9530, USA}

\author{A.N. Cleland}
\affiliation{Department of Physics, University of California, Santa Barbara, California 93106-9530, USA}
\affiliation{California NanoSystems Institute, University of California, Santa Barbara, CA 93106-9530, USA}
\author{John M. Martinis}
\email{martinis@physics.ucsb.edu}
\affiliation{Department of Physics, University of California, Santa Barbara, California 93106-9530, USA}
\affiliation{California NanoSystems Institute, University of California, Santa Barbara, CA 93106-9530, USA}

\date{\today}
\begin{abstract}
Superconducting microwave circuits based on coplanar waveguides (CPW) are susceptible to parasitic slotline modes which can lead to loss and decoherence. We motivate the use of superconducting airbridges as a reliable method for preventing the propagation of these modes. We describe the fabrication of these airbridges on superconducting resonators, which we use to measure the loss due to placing airbridges over CPW lines. We find that the additional loss at single photon levels is small, and decreases at higher drive powers.
\end{abstract}
\maketitle

Superconducting coplanar waveguide (CPW) transmission lines and resonators are integral components of cryogenic detectors for submillimeter electromagnetic radiation, \cite{mazin2012superconducting,day2003broadband} quantum memory elements, \cite{hofheinz2009synthesizing} and solid state quantum computing architectures. \cite{galiautdinov2012resonator,mariantoni2011implementing,barends2013coherent} The desired mode profile of a CPW is symmetric,\cite{simons2004coplanar} with the two ground planes on either side of the center trace held to the same voltage. However, asymmetries and discontinuities in the microwave circuitry can lead to the excitation of parasitic slotline modes.\cite{ponchak2005excitation} These modes can couple to elements of the circuit such as qubits, and they represent a source of radiation loss and decoherence.\cite{harokopus1991radiation,houck2008controlling} 
	
In order to suppress these spurious modes, crossover connections need to be made between the ground planes that are interrupted by the CPW structure. Free standing crossovers, known as airbridges, have been a staple of conventional microwave CPW technology, \cite{koster1989investigations,kwon2001low} and fabrication processes have recently been developed for building airbridges on superconducting microwave circuits. \cite{lankwarden2012development,abuwasib2013fabrication} However, the fabrication of airbridges adds additional processing that may degrade the quality of the circuit, and the airbridges themselves may present a source of loss. In addition, care must be taken in order to avoid accidentally creating tunnel junctions with small critical currents at the interfaces of such structures. In this Letter, we present the first characterizations of the loss due to fabricating airbridge crossovers on superconducting microwave resonators. We find that the loss due to airbridge crossovers is small but not negligible, and should be taken into account when engineering crossovers for low loss circuit elements.

\begin{figure}
\begin{centering}
\includegraphics{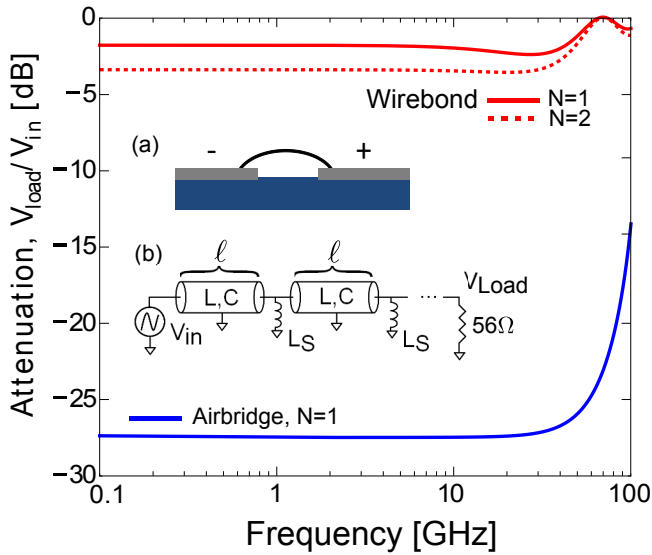} 
\par\end{centering}
\caption{(Color online): (Inset) (a) The slotline mode of a CPW is modeled by removing the center trace. A crossover wire is used to tie together the two planes so that the slotline mode may not propagate. (b) Equivalent transmission line model for the slotline mode shunted by crossovers with an inductance $L_{S}$. We obtained a capacitance and inductance per length of C$=$140\,fF/mm and L$=$450\,pH/m from numerical simulation of a 20\,$\mu$m gap slotline, giving an impedance of 56\,$\Omega$ which is matched by the load. The wirebond and airbridge have an $L_{S}$ of 1\,nH and 10\,pH respectively, and are placed at intervals of length $\ell$. Main panel: SPICE simulations for 1\,mm of the transmission line model, showing that the attenuation due to a single airbridge is more than 20\,dB greater than for a wirebond. Ten airbridges per mm can be simply fabricated and gives an attenuation of -150\,dB (not shown)}
\label{figure:theory} 
\end{figure}

To motivate our use of airbirdges, we observe that in past work with superconducting circuits, connections between the different ground planes have been typically been made using wirebonds. However, with a wire diameter of 1\,mil and a typical length of 1\,mm, wirebonds have an inductance of order 1\,nH \cite{rosabulletin} and an impedance ~40\,$\Omega$ at 6\,GHz, making them an ineffective shunt. In comparison, airbridges have 100 times less inductance due to their small size. In order to understand the effect of the crossover impedance on slotline attenuation, we studied a simple transmission line model \cite{wenner2011wirebond} for the slotline with evenly spaced inductive shunts to ground as shown in Fig. 1(a). We simulated in SPICE 1\,mm of a transmission line with with a terminated load, and varied the number of inductive shunts. As seen in Fig. 1, the attenuation per millimeter of the slotline propagation for a single airbridge is two orders of magnitude greater than for one or two wirebonds. This result can be easily understood if we consider only the inductances of the model, which is valid below the cutoff frequency.\cite{wenner2011wirebond} The inductance of 0.5\,mm of the slotline is 0.23\,nH, which is smaller than the wirebond inductance but much bigger than the inductance of an airbridge. Thus, in the case of wirebonds, signal will continue propagating down the line rather than flow to ground. Furthermore, while increasing the wirebond density can be difficult and unreliable, increasing the airbridge density can be done by simply redesigning a photomask. With 10 airbridges per mm, we simulated the attenuation to be -150\,dB, implying the slotline mode does not exist.

\begin{figure}
\begin{centering}
\includegraphics{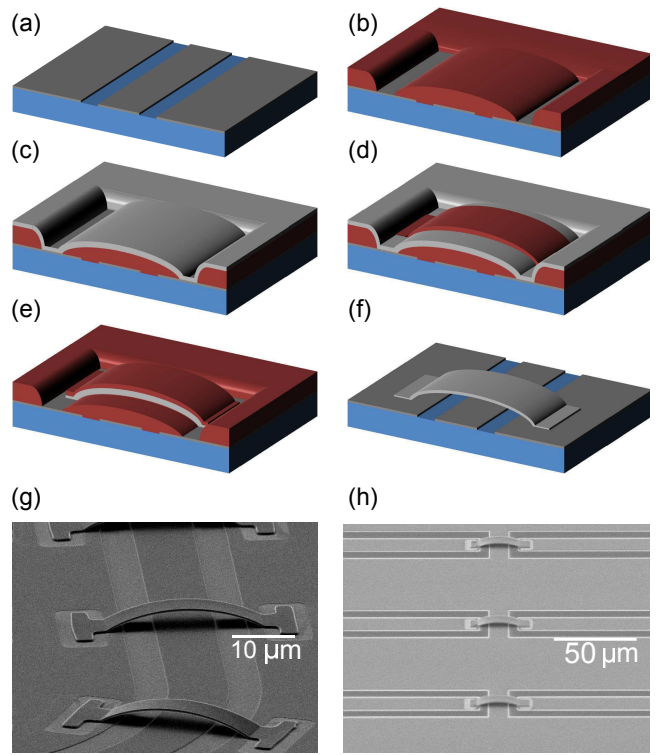} 
\par\end{centering}
\caption{(Color online) (a-f) Fabrication process for superconducting airbridges, with substrate shown in blue, resist in dark red and aluminum in gray. In order, the fabrication steps are: (a) fabrication of CPW base layer, (b) patterning and reflow of photoresist, (c) deposition of aluminum, (d) definition of the bridge using lithography, (e) wet etching of excess aluminum and, (f) release of airbridge. (g) SEM image of airbridges connecting the ground planes of a CPW line and (h) SEM image of airbridges linking together two CPW lines.}
\label{figure:fab} 
\end{figure}
The fabrication process we used for the airbridges follows from earlier work done on kinetic inductance detectors,\cite{lankwarden2012development} with modifications to adapt the process for an aluminum base layer. We show the process flow and resulting structures in Fig. 2. First, we formed the scaffold for the bridge from a 3\,$\mu$m thick positive photoresist (Megaposit SPR-220-3). The thickness of the bridge is set by the resist thickness, and photolithography determines the placement and span of the bridge. Throughout the process we used a developer (AZ Dev 1:1) designed to minimize aluminum etching. Away from the bridge area, we did not expose the resist so that it remained as a protective layer and etch stop. Next, we reflowed the resist at 140$^\circ$C for 3 minutes to form an arch for mechanical stability. We then deposited 300\,nm of aluminum in a high vacuum electron beam evaporator to form the bridge layer. Prior to the deposition, we used an in-situ argon ion mill calibrated to remove the native oxide of the base aluminum in order for the bridges to make good electrical contact.\cite{barends2013coherent} The ion mill was operated for 3.5 minutes in 1$\times 10^{-4}$\,mbar of argon, with beam voltage 400V, beam current 21\,mA, and beam width 3.2".  Using a second layer of patterend 3\,$\mu$m resist as a mask, we then wet etched (Transene Aluminum Etchant Type A at 30$^\circ$C) the excess deposited aluminum that is not used to form the bridge. We determined the termination of the etch by visual inspection. When the top layer of aluminum was etched away, the wafer went through a clear change in reflectance from aluminum to the underlying resist layer; the typical etch time was 5 minutes. We continued immersing the wafer in the etchant for 5 seconds after this transition, then immersed the wafer in water for 3 minutes. The termination of the etch is a critical step because the regions around the pads of the bridge are not protected by photoresist during the etch, and can potentially be etched through, breaking the ground plane. Finally, we stripped both layers of resist in an 80$^\circ$C heated bath of Microposit 1165 photoresist stripper to release the airbridges.

We initially confirmed that the argon ion mill step led to an ohmic contact at the bridge pads by measuring the room temperature resistance across multiple airbridges using a standard four terminal measurement. For ten airbridges in series that were 28 microns long and 8 microns wide, we found the resistance to be 6\,$\Omega$, which is roughly consistent with the resistance of an aluminum wire of the same geometry. We further ruled out any thick oxide junctions by measuring the critical current across multiple airbridges below the critical temperature (1.2\,K), which we found to be of order 100\,mA.\cite{supplement}

In developing the process, we initially found a large amount of residual resist remained from the scaffolding layer after stripping in solvents. This decreased the reliability of our bridges by loading and deforming the bridge arches, and would have contributed a large amount of loss to our circuit. We deduced that the residue consisted of resist  cross-linked by ion implantation from the argon ion mill step, a well known problem in semiconductor processing.\cite{okuyama1978high} We were able to mitigate the problem by stripping the resist layers in a downstream oxygen plasma at 150$^\circ$C for 5 minutes prior to stripping in a solvent bath. The low temperature oxygen plasma acts to burn off the damaged layer of resist. 

With this additional cleaning step, we have been able to reliably fabricate bridges over a range of spans from 2\,$\mu$m to 50\,$\mu$m. The main sources of bridge failure are factors other than their structural stability such as lithographic errors, and the failure rate is less than 0.1\%. We have also tested the bridges in a variety of postprocessing steps, including wafer dicing and fabricating aluminum junctions with a bilayer electron beam resist process; bridges spanning up to 40\,$\mu$m have been found to survive these steps reliably. We note here that the airbridges generally do not survive sonication.

\begin{figure}
\begin{centering}
\includegraphics{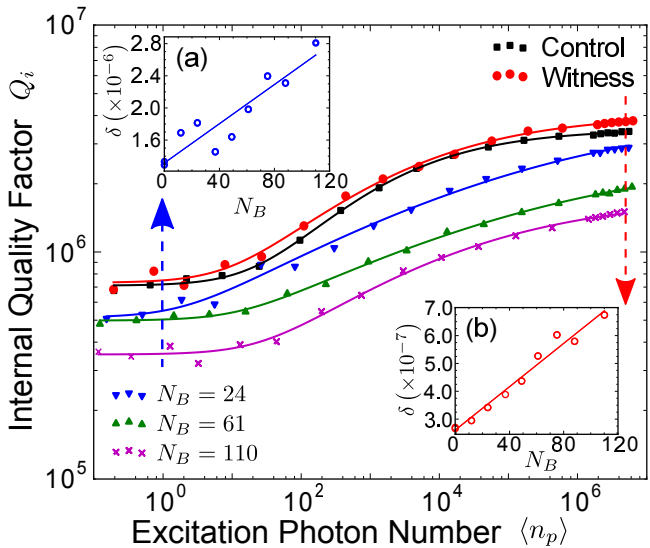} 
\par\end{centering}
\caption{(Color online) Main panel: Dependence of the internal quality factor $Q_{i}$ on the average photon population in the resonator $\langle n_{p} \rangle$, shown for various numbers of airbridges $N_{B}$ spanning the ground planes of the resonator. Data for a control resonator and a witness resonator are also displayed. Lines are guides for the eye. Insets: (a) Total loss tangent $\delta$ at single photon  as a function of number of airbridges. The best fit line has slope $1.2 \times 10^{-8}$ . (b) Loss at $\langle n_{p} \rangle=5 \times 10^{6}$, with slope $3.8 \times 10^{-9}$.}
\label{figure:loss} 
\end{figure}
To measure the loss added by placing an airbridge over a CPW transmission line, we constructed quarter wavelength CPW resonators with variable numbers of airbridges connecting the ground planes. We fabricated the resonators from an aluminum film deposited on a high-resistivity silicon subtrate and etched with a BCl$_{3}$/Cl$_{2}$ inductively coupled plasma.\cite{megrant2012planar} We have also used the process on a sapphire substrate with comparable results. We designed the resonators with 10\,$\mu$m center traces and 5\,$\mu$m gaps to match the dimensions of our typical feedlines, and designed the resonant frequencies to range from 5 to 6\,GHz. We designed the airbridges to have 4\,$\mu$m of clearance from the CPW line for a total length of 28\,$\mu$m, and chose an airbridge width of 8\,$\mu$m to ensure mechanical stability of the bridge. On eight of the resonators, we fabricated between 12 and 110 airbridges spanning the resonator center trace, evenly spaced in the number of bridges. The resonators with the most airbridges had a density of one airbridge every 50\,$\mu$m, covering 16\% of the total length of the resonator. The remaining two witness resonators went through the full process of airbridge fabrication but were not designed with any airbridges spanning their center traces. We used these witness resonators as a test of whether placing airbridges on a CPW line adds loss to other circuit elements on the same wafer. We also fabricated a separate chip of resonators from the same film that saw no additional processing, to act as a control sample.

In order to determine the internal quality factor $Q_{i}$ of the resonators, we placed them in an adiabatic demagnetization refrigerator (ADR), which reached a base temperature of 50\,mK. We determined $Q_{i}$ by measuring the transmission through a feedline that was capacitively coupled to each resonator (see Ref. \onlinecite{megrant2012planar} for measurement details). We varied the drive power such that the photon population $\langle n_p \rangle$ in the resonator ranged from single photon levels up to $10^7$ photons, at which point the resonators became non-linear. A sample of representative quality factor data for some of the resonators is shown in the main panel of Fig. 3. As seen in previous work, the quality factor of the resonators increases as a function of increasing drive power, which is consistent with the loss in the resonator being dominated by two-level states (TLS) at the material interfaces.\cite{megrant2012planar,gao2008experimental,martinis2005decoherence,sage2011study}

In general, we expect the dependence of the quality factor to show two plateaus, one around single photon levels corresponding to loss being dominated by TLS, and one at high power corresponding to saturation of TLS.\cite{megrant2012planar,sage2011study} From our control chip, for which a representative example is shown in black squares in Fig. 3, we determined that the nominal internal quality factor for our resonator geometry and material was around $7.0 \times 10^5$ at single photon powers and $3.5 \times 10^6$ at high power (5 million photons). Our witness resonators, for which a representative example is shown in red circles in Fig. 3, performed similarly, indicating that the additional processing on the chip did not add any loss. As we increased the number of bridges fabricated over the resonator, the quality factor decreased at both low and high powers. Interestingly, the quality factor does not appear to plateau as strongly at high power when airbridges are present. 

To determine quantitatively the loss due to airbridges, we extracted the loss tangent $\delta=1/Q_{i}$ at powers around a single photon and at 5 million photons. In Fig. 3a (3b), we plot the low power (high power) loss tangent as a function of number of airbridges, along with lines of best fit. From the slopes, we estimate that each airbridge adds $1.2 \times 10^{-8}$ to the loss tangent of the resonator at low power and $3.8 \times 10^{-9}$ at high power. We can also estimate the loss per fraction of the resonator that is covered by the resonator. If we assume that the loss also scales with the width of the airbridges, then every one percent of the resonator covered by airbridges adds an additional $8.3 \times 10^{-8}$ to the loss tangent at low power and $2.7 \times 10^{-8}$ at high power. We note that a resonator completely covered by airbridges would be limited to an internal quality factor of order 120,000 at low power, which is more than five times lower than the uncovered device.

To understand this increase in the loss due to airbridges, we note that the change in the loss at different drive powers suggests that the addition of an airbridge adds to the TLS loss of the resonator. However, the source of this additional TLS loss is unclear. Optical and SEM images show that despite our efforts to clean off the resist, residues still remain on some edges close to the pads of the bridge, some of which is visible in the lower left and top right contact pads shown in Fig. 2(g). The interface underneath the bridge is also suspect, since it was deposited on photoresist that had been crosslinked by the argon ion mill. However, the participation ratios of these regions are small since they are not in the regions of high field.\cite{wenner2011surface} As an example, through a simulation in COMSOL, we find that adding 100 nm of dielectric material with a loss tangent of $10^{-3}$ (typical for amorphous oxides) and dielectric constant of 4 gives a loss tangent of $1.6 \times 10^{-6}$, or a $Q_{i}$ of 630,000 for complete airbridge coverage (see supplement for details). This is significantly higher than our measured value despite the fact that 100\,nm is the upper bound on the interface thickness under the bridge based on SEM images.\cite{supplement}

In addition to placing airbridges over CPW lines to connect the ground planes together, we have also fabricated airbridges to connect two CPW center traces together. We tested such a connection by fabricating quarter wavelength CPW resonators with intentional breaks in the resonator, then reconnecting the lines with airbridges, as shown in Fig. 2(h). The resonators fabricated using this method performed comparably to resonators fabricated with ground plane airbridges, with each airbridge connection added to the center trace adding a loss of $1 \times 10^{-7}$. More information about these measurements can be found in the supplement. These results indicate that airbridges can used be to carry microwave signals, and for example, allowing for the crossing of two perpendicular superconducting CPW lines.\cite{supplement}

In summary, we have fabricated reliable aluminum airbridges on an aluminum microwave circuit. Simple considerations of the inductances of a wirebond versus an airbridge show that airbridges are far superior for the purpose of attenuating slotline modes. However, placing them on CPW lines adds some additional loss proportional to the amount of airbridge coverage, which should be taken into consideration when building sensitive, low-loss microwave elements. 

The authors thank J. Baselmans for helpful discussions and for providing test samples. Devices were fabricated at the UCSB Nanofabrication Facility, a part of the NSF-funded National Nanotechnology Infrastructure Network, and at the NanoStructures Cleanroom Facility.  This research was funded by the Office of the Director of National Intelligence (ODNI), Intelligence Advanced Research Projects Activity (IARPA), through Army Research Office Grant No. W911NF-09-1-0375. All statements of fact, opinion, or conclusions contained herein are those of the authors and should
not be construed as representing the official views or policies
of IARPA, the ODNI, or the U.S. Government.

\end{document}


\author{Zijun Chen}
\affiliation{Department of Physics, University of California, Santa Barbara, California 93106-9530, USA}
\author{A. Megrant}
\affiliation{Department of Physics, University of California, Santa Barbara, California 93106-9530, USA}
\affiliation{Department of Materials, University of California, Santa Barbara, California 93106, USA}
\author{J. Kelly}
\affiliation{Department of Physics, University of California, Santa Barbara, California 93106-9530, USA}
\author{R. Barends}
\affiliation{Department of Physics, University of California, Santa Barbara, California 93106-9530, USA}
\author{J. Bochmann}
\affiliation{Department of Physics, University of California, Santa Barbara, California 93106-9530, USA}
\author{Yu Chen}
\affiliation{Department of Physics, University of California, Santa Barbara, California 93106-9530, USA}
\author{B. Chiaro}
\affiliation{Department of Physics, University of California, Santa Barbara, California 93106-9530, USA}
\author{A. Dunsworth}
\affiliation{Department of Physics, University of California, Santa Barbara, California 93106-9530, USA}
\author{E. Jeffery}
\affiliation{Department of Physics, University of California, Santa Barbara, California 93106-9530, USA}
\author{J.Y. Mutus}
\affiliation{Department of Physics, University of California, Santa Barbara, California 93106-9530, USA}
\author{P.J.J. O'Malley}
\affiliation{Department of Physics, University of California, Santa Barbara, California 93106-9530, USA}
\author{C. Neill}
\affiliation{Department of Physics, University of California, Santa Barbara, California 93106-9530, USA}
\author{P. Roushan}
\affiliation{Department of Physics, University of California, Santa Barbara, California 93106-9530, USA}
\author{D. Sank}
\affiliation{Department of Physics, University of California, Santa Barbara, California 93106-9530, USA}
\author{A. Vainsencher}
\affiliation{Department of Physics, University of California, Santa Barbara, California 93106-9530, USA}
\author{J. Wenner}
\affiliation{Department of Physics, University of California, Santa Barbara, California 93106-9530, USA}
\author{T.C. White}
\affiliation{Department of Physics, University of California, Santa Barbara, California 93106-9530, USA}
\author{A.N. Cleland}
\affiliation{Department of Physics, University of California, Santa Barbara, California 93106-9530, USA}
\affiliation{California NanoSystems Institute, University of California, Santa Barbara, CA 93106-9530, USA}
\author{John M. Martinis}
\email{martinis@physics.ucsb.edu}
\affiliation{Department of Physics, University of California, Santa Barbara, California 93106-9530, USA}
\affiliation{California NanoSystems Institute, University of California, Santa Barbara, CA 93106-9530, USA}
\title{Supplementary Material for "Fabrication and Characterization of Aluminum Airbridges for Superconducting Microwave Circuits"}

\date{\today}

\begin{abstract}

We provide supplementary data and calculations, primarily for practical design considerations when building superconducting circuits using airbridges. We also show a sample calculation of the contribution of an airbridge to microwave loss.

\end{abstract}

\maketitle

\section{Estimation of Airbridge Critical Current}
\begin{figure}
\begin{centering}
\includegraphics{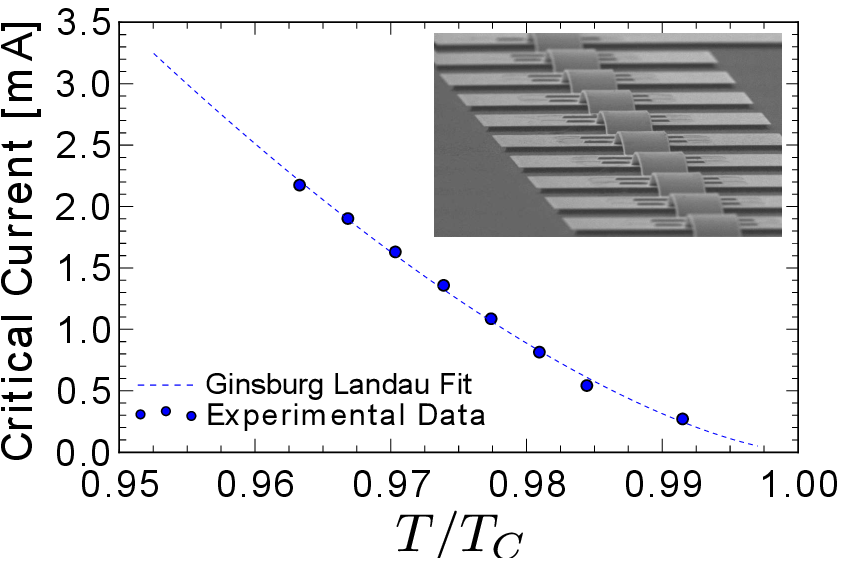} 
\par\end{centering}
\caption{(Color online): Inset: Ten airbridges fabricated in series for a four terminal measurement of the resistance. Main panel: Critical current as a function of reduced temperature $T/T_{c}$. The fit is to Eqn. 1, with $I_0=462\, mA$}
\label{figure:IV} 
\end{figure}

As detailed in the main paper, we fabricated 10 airbridges in series and measured them in a four terminal configuration. Each airbridge had a width of 8\,$\mu$m, a length of 28\,$\mu$m, and a thickness of 300\,nm. At room temperature, we measured a resistance of 6\,$\Omega$. For a standard aluminum resistivity of 2.7$\times 10^{-6} \, \Omega$-cm, the expected resistance at room temperature is 3.15 $\Omega$, which does not take into account the curvature of the bridges and the distance between the pads of the bridges, which was 6 microns. At 100 mK, we were limited to 10 mA of drive current, which was not enough to drive the airbridges normal. Instead, we slowly cooled the sample through $T_{c}=1.2K$ and measured the critical current $I_c$ as a function of temperature just below $T_{c}$, with the results shown in Fig. 1. The data matches the expected Ginsburg-Landau behavior, which predicts the following relation for the critical current of a thin superconducting wire \cite{tinkham2004introduction}
\begin{equation}
I_c=I_0 \left(1-T/T_c \right)^{2/3}
\end{equation}
where $I_0$ is the critical current at temperature well below $T_c$. By fitting to this equation, we extracted a low temperature critical current of 462\, mA. However, this result does not take into account the width of our airbridges. From previous works, we estimate that there is a decrease in $I_0$ by a factor of order 3 or 4 for an 8\,$\mu$m wire,\cite{skocpol1976critical} giving a critical current of around 100\,mA.

\section{Shifts in Resonant Frequency Due to Airbridges}
\begin{figure}
\begin{centering}
\includegraphics{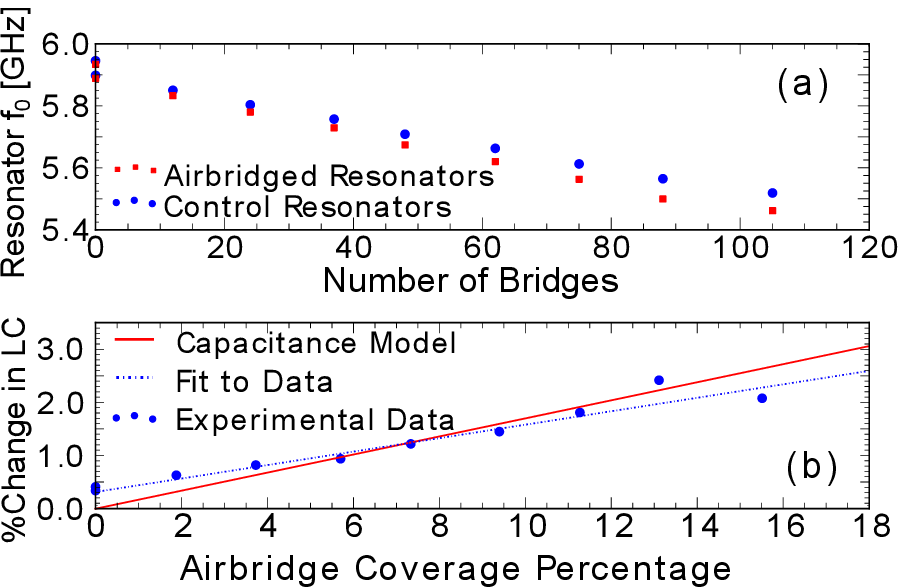} 
\par\end{centering}
\caption{(Color online): (a) Resonant frequencies for resonators with variable numbers of airbridges in red squares, compared with the frequencies of their corresponding controls which are designed to have the same length. As the number of bridges increases, the resonators shift lower in frequency compared to their controls. (b) Percent change in $LC$, the product of the inductance per length and capacitance per length, as a function of the percentage of the resonator covered by airbridges. The dashed blue line is a linear fit to the data, with slope 12.7\% and intercept 0.35\%. The offset from the origin is within normal chip to chip variations in our measured resonators. The red line is a prediction based on the additional capacitance of the airbridge. The slopes differ due to the decrease in inductance from the airbridge.}
\label{figure:LC} 
\end{figure}

Compared to more conventional crossovers which are supported by dielectrics, airbridges have a much smaller impact on the capacitance of a CPW line. However, this additional capacitance due to an airbridge is not negligible and should be accounted for. For example, in our experiment to test the microwave loss of airbridges using ten different resonators, we designed the resonators such that the density of airbridges increased with decreasing frequency, as shown in Fig. 2(a). A higher density of airbridges increases the capacitance of the resonator and decreases the resonant frequency. Thus, in our experiment, the resonant frequencies shifted further apart rather than closer together, avoiding any frequency collisions. We note here that from our control data, we found no significant correlation of the high or low power quality factor with the frequency of the resonator over the range we considered, which validated this particular design choice.

If we assume the airbridge acts like a parallel plate capacitor between the center trace and ground, we can estimate the additional capacitance per unit length due to the airbridge as $C=\epsilon_0 w/d$, where $w$ is the width of the center trace and $d$ is the height of the airbridge. For the geometry in our experiment, $w=10 \mu$m and $d=3 \mu$m, giving $C=29.5$ pF/m. We can also numerically calculate the additional capacitance due to the airbridge using COMSOL. We simulated the cross-section of a CPW line with a 10 $\mu$m center trace and 5 $\mu$m gap with a substrate dielectric constant of 11.6, and found the capacitance per length to be 175.25 pF/m. After adding an airbridge, the capacitance increased to 204.03 pF/m giving an increase of 28.78 pF/m due to the airbridge, showing remarkable agreement with the parallel plate estimate. From these values, we predict that the capacitance of a resonator covered completely by airbridges should increase by 17\%.

From the frequency data shown in Fig. 2(a), we can determine the actual effect of placing an airbridge over a CPW line. As the number of airbridges increased, the frequency of the resonator shifted further below the frequency of its corresponding control. Since each resonator and its control are designed to have the same wavelength, we can interpret the change in frequency as a change in the phase velocity of light $v_p=1/\sqrt{LC}$, where $L$ and $C$ are the inductance and capacitance per unit length. Given the total length of the resonator and the number of airbridges, we can also determine the percentage of the line covered by airbridges. The percent coverage should be linearly related to the change in the product of the inductance and capacitance per unit length, $LC$, which is shown in Fig. 2(b). The slope of the linear fit in Fig. 2(b) indicates that the $LC$ product for a section of line covered by airbridge differs from the bare line by 12.7\%.

The discrepancy between our prediction and our data is most likely due to changes in the inductance of the resonator. Each airbridge adds additional pathways for current to flow, which decreases the inductance of the CPW line and compensates in part for the increase in capacitance. However, the inductance is not as easily modeled as the capacitance since edge effects are important. In other words, a single, wide airbridge that spanning a CPW line does not have the same effect as multiple narrower airbridges because they contain different current paths.

\section{Participation Ratio of the Airbridge Interface}
\begin{figure}
\begin{centering}
\includegraphics{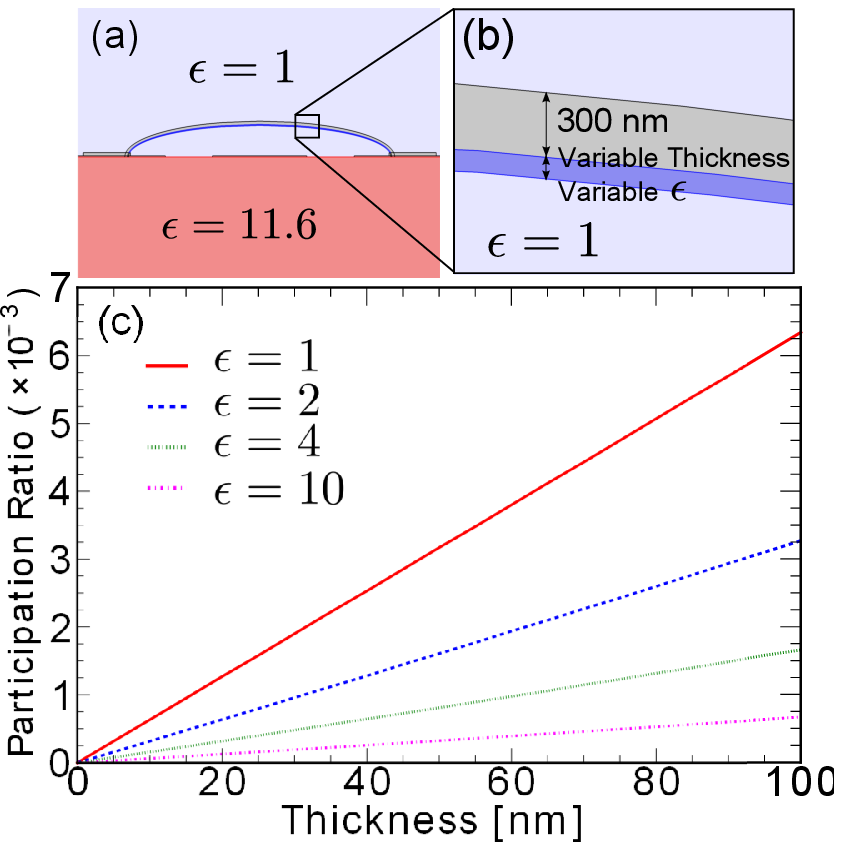} 
\par\end{centering}
\caption{(Color online): (a) Cross section of an airbridge spanning a CPW line. (b) Close-up of the interface for which we calculate the participation ratio. This interface is a possible source of loss because the layer of aluminum is deposited on photoresist that has been crosslinked by an ion mill. The thickness and dielectric constant are variable. (c) Participation ratio as a function of thickness for various dielectric constants at the interface. We numerically calculate using COMSOL the participation ratio by setting the potential of the center trace of the CPW to 1V, solving for the electric fields, then numerically computing the integral in Eq. 1 in the interface region. We obtain the total energy $W$ by performing the same integral for all of the cross section.}
\label{figure:participation} 
\end{figure}
The interface underneath the airbridge is a potential source of loss, since this is the interface at which we deposited aluminum on photoresist that has been crosslinked by the argon ion mill. To understand the additional surface loss due to this interface, we calculate the participation ratio of a lossy dielectric at this metal-air interface following Ref \onlinecite{wenner2011surface}. We consider the resonator and airbridge structure in cross section as shown in Fig. 3(a). The participation ratio $p$ of any isotropic region of space in this cross-section is simply given by the ratio of energy stored in the region to the total energy stored in the entire cross-section
\begin{equation}
p=W^{-1} \epsilon_r \epsilon_0 \int \!\!\! \int dA \,\frac{\vert E \vert ^2}{2}
\end{equation}
where $W$ is the total energy in the cross-section which may be obtained by performing the same integral over all space, and $\epsilon_r$ is the dielectric constant in the region. Assuming that the region is thin, as it is in the case of our interface of interest, we can replace an integral over the thickness by a product, turning the double integral into a line integral over the boundary of the interface. We can also simplify the equation using the boundary conditions on our interface. The metal boundary allows us to approximate the electric field as normal to the metal, while the continuity of the displacement field at the air interface gives us the relation $\epsilon_r E_{i \perp}=E_{a \perp}$, where $E_{i}$ is the electric field in the interface and $E_{a}$ is the electric field in air. Combining these simplifications we obtain
\begin{equation}
p=W^{-1} t_{i} \epsilon_r ^{-1} \epsilon_0 \int dS \,\frac{\vert E_{a \perp} \vert ^2}{2}
\end{equation}
where $t_{i}$ is the small thickness of the interface. Assuming the contribution to the total energy $W$ of the interface is small, the participation ratio is proportional to the thickness and inversely proportional to the dielectric constant. We can estimate the value of the line integral by again modeling the airbridge as a parallel plate. If we assume a 1\,V difference in potential between center trace and ground, then from the calculation of total capacitance above, we know the value of $W=\frac{1}{2}C V^2$. The electric field is given by 1\,V divided by the separation distance of 3 $\mu$m, and we may replace the integral with a multiplication by the length, about 10 $\mu$m. We then obtain the following approximate formula:
\begin{equation}
p=4.8 \times 10^{-5}\,\textrm{nm}^{-1}\, \frac{t_i}{\epsilon_r}
\end{equation}
Alternatively, we can also numerically evaluate Eq. 1. We constructed the geometry of an airbridge spanning a CPW and included a thin dielectric interface on the underside of the bridge as shown in Fig. 4(b). After applying a potential of 1\,V to the center trace, we solved for the electric fields and numerically integrated Eq. 1 to determine the total energy in the cross-section and the energy in the interface, giving us the participation ratio. We calculated participation ratio as a function of interface thickness and dielectric constant, producing the plot shown in Fig. 4(c). We see that the scaling follows the expected scaling from Eqs. 2 and 3. Furthermore, we can more accurately determine the coefficient in Eq. 3 from the slopes of the lines, and we find that the coefficient is $6.34 \times 10^{-5}\,nm^{-1}$, which is within 30\% of our parallel plate estimation.

Given the participation ratio, we can estimate the loss due to this interface. For the dielectric constant, we estimate a dielectric constant of 4 based on data pertaining to other photoresists \cite{pierce1965dielectric}. SEM images of the interface were inconclusive for determining the thickness, but it is certainly upper bounded by 100\,nm. Finally, there is little data on the loss tangent of resist and cryogenic temperatures, so we estimate this to be $10^{-3}$ based on the measured loss tangents of amorphous oxides.\cite{oconnell2008microwave} Using these numbers, we obtain a participation ratio of $1.6 \times 10^{-3}$ and a loss due to airbridges of $1.6 \times 10^{-6}$, or a $Q_i$ of 630,000.

\section{Loss due to Inline Airbridges}
\begin{figure}
\begin{centering}
\includegraphics{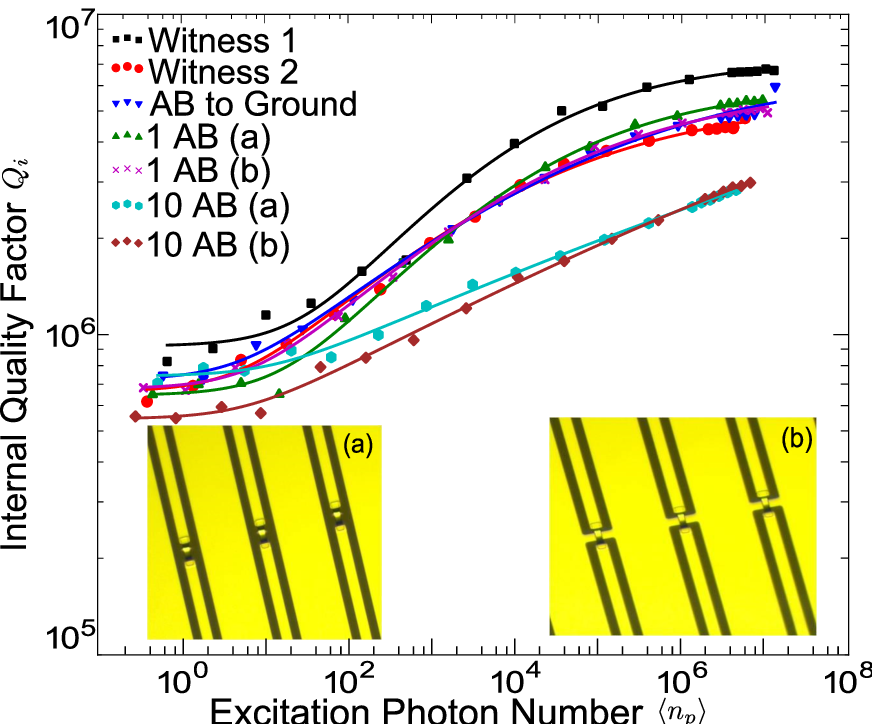} 
\par\end{centering}
\caption{(Color online): Insets: (a) Airbridges connecting together CPW lines within a resonator. (b) A second style of airbridge connection, where the ground plane is threaded underneath the airbridge. Main panel: Internal quality factor of resonators as a function of average photon population for many different styles of resonators. We show two witness resonators to demonstrate the typical spread in measured $Q_{i}$. Lines are guides for the eye.}
\label{figure:CTAB} 
\end{figure}

Given the high critical currents through the airbridges, we know that the airbridges provide a good connection at DC. In order to test connectivity at microwave frequencies, we fabricated airbridges as a part of the center trace of the quarter wave resonators described in the main paper. We considered two styles of inline airbridges. In both styles, we design the center trace to have a 20\,$\mu$m break, then connect together the two traces with an airbridge. In one style shown in Figure 4(a), the ground plane is left unconnected, while in the other style shown in Fig. 4(b), the ground plane is connected through the break in the center trace and underneath the bridge.

We tested one and ten inline airbridges placed inside quarter wave CPW resonators in both styles. In addition, we tested a quarter wave resonator with an inline airbridge acting as the short to ground, since this configuration gave the largest current loading of the airbridge. Based on loss results in the main paper, we were confident that the airbridge processing did not degrade the quality factors of our resonators and used witness resonators fabricated on the same chip as the control resonators. All resonators had a larger center trace of 15\,$\mu$m  to accommodate the pads of the bridge, a gap of 10\,$\mu$m , and were fabricated using aluminum deposited on a sapphire substrate. We performed quality factor measurements as described in the main paper, producing the results shown in Fig. 4.

The two witness resonators shown in Fig. 4 represent the best and worst measured quality factors for our witness resonators. On average, the witness resonators show a low power $Q_{i}$ of around 800,000, and a high power $Q_{i}$ of around $5 \times 10^6$. All resonators which have a single inline airbridge, including the resonator shorted to ground by the airbridge, do not show substantial degradation in $Q_{i}$. On the other hand, ten inline airbridges shows some degradation at high power corresponding to a additional loss of $3 \times 10^{-7}$. Ten inline airbridges with threaded ground planes also showed significant loss at lower power, with an additional loss of $1.3 \times 10^{-6}$, or $10^{-7}$ per bridge.